\documentstyle{mn}
\newread\epsffilein    
\newif\ifepsffileok    
\newif\ifepsfbbfound   
\newif\ifepsfverbose   
\newdimen\epsfxsize    
\newdimen\epsfysize    
\newdimen\epsftsize    
\newdimen\epsfrsize    
\newdimen\epsftmp      
\newdimen\pspoints     
\def\epsfbox#1{\global\def\epsfllx{72}\global\def\epsflly{72}%
   \global\def\epsfurx{540}\global\def\epsfury{720}%
   \def\lbracket{[}\def\testit{#1}\ifx\testit\lbracket
   \let\next=\epsfgetlitbb\else\let\next=\epsfnormal\fi\next{#1}}%
\def\epsfgetlitbb#1#2 #3 #4 #5]#6{\epsfgrab #2 #3 #4 #5 .\\%
   \epsfsetgraph{#6}}%
\def\epsfnormal#1{\epsfgetbb{#1}\epsfsetgraph{#1}}%
\def\epsfgetbb#1{%
\openin\epsffilein=#1
\ifeof\epsffilein\errmessage{I couldn't open #1, will ignore it}\else
   {\epsffileoktrue \chardef\other=12
    \def\do##1{\catcode`##1=\other}\dospecials \catcode`\ =10
    \loop
       \read\epsffilein to \epsffileline
       \ifeof\epsffilein\epsffileokfalse\else
          \expandafter\epsfaux\epsffileline:. \\%
       \fi
   \ifepsffileok\repeat
   \ifepsfbbfound\else
    \ifepsfverbose\message{No bounding box comment in #1; using defaults}\fi\fi
   }\closein\epsffilein\fi}%
\def\epsfsetgraph#1{%
   \epsfrsize=\epsfury\pspoints
   \advance\epsfrsize by-\epsflly\pspoints
   \epsftsize=\epsfurx\pspoints
   \advance\epsftsize by-\epsfllx\pspoints
   \epsfxsize\epsfsize\epsftsize\epsfrsize
   \ifnum\epsfxsize=0 \ifnum\epsfysize=0
      \epsfxsize=\epsftsize \epsfysize=\epsfrsize
     \else\epsftmp=\epsftsize \divide\epsftmp\epsfrsize
       \epsfxsize=\epsfysize \multiply\epsfxsize\epsftmp
       \multiply\epsftmp\epsfrsize \advance\epsftsize-\epsftmp
       \epsftmp=\epsfysize
       \loop \advance\epsftsize\epsftsize \divide\epsftmp 2
       \ifnum\epsftmp>0
          \ifnum\epsftsize<\epsfrsize\else
             \advance\epsftsize-\epsfrsize \advance\epsfxsize\epsftmp \fi
       \repeat
     \fi
   \else\epsftmp=\epsfrsize \divide\epsftmp\epsftsize
     \epsfysize=\epsfxsize \multiply\epsfysize\epsftmp   
     \multiply\epsftmp\epsftsize \advance\epsfrsize-\epsftmp
     \epsftmp=\epsfxsize
     \loop \advance\epsfrsize\epsfrsize \divide\epsftmp 2
     \ifnum\epsftmp>0
        \ifnum\epsfrsize<\epsftsize\else
           \advance\epsfrsize-\epsftsize \advance\epsfysize\epsftmp \fi
     \repeat     
   \fi
   \ifepsfverbose\message{#1: width=\the\epsfxsize, height=\the\epsfysize}\fi
   \epsftmp=10\epsfxsize \divide\epsftmp\pspoints
   \newcount\figskipcount
      \message{#1 \the\epsfysize  }
   \vbox to\epsfysize{\vfil\hbox to\epsfxsize{%
      \includegraphics{#1}%
      \hfil}}%
\epsfxsize=0pt\epsfysize=0pt}%

{\catcode`\%=12 \global\let\epsfpercent=
\long\def\epsfaux#1#2:#3\\{\ifx#1\epsfpercent
   \def\testit{#2}\ifx\testit\epsfbblit
      \epsfgrab #3 . . . \\%
      \epsffileokfalse
      \global\epsfbbfoundtrue
   \fi\else\ifx#1\par\else\epsffileokfalse\fi\fi}%
\def\epsfgrab #1 #2 #3 #4 #5\\{%
   \global\def\epsfllx{#1}\ifx\epsfllx\empty
      \epsfgrab #2 #3 #4 #5 .\\\else
   \global\def\epsflly{#2}%
   \global\def\epsfurx{#3}\global\def\epsfury{#4}\fi}%
\def\epsfsize#1#2{\epsfxsize}
\def\figinsert#1#2{\epsfbox{#1} \message{#2} }    
\title[The Durham/UKST Galaxy Redshift Survey]
      {The Durham/UKST Galaxy Redshift Survey - III.\\ Large Scale
Structure via the 2-Point Correlation Function.}
\author[A. Ratcliffe et al.]
       {A. Ratcliffe$^{1}$, T. Shanks$^{1}$, Q.A. Parker$^{2}$ and R.
Fong$^{1}$ \\
        $^{1}$Physics Deptartment, University of Durham, South Road, Durham,
DH1 3LE.\\
	$^{2}$Anglo-Australian Observatory, Coonabarabran, NSW 2357,
Australia.}
\begin{document}

\maketitle

\begin{abstract}
We have investigated the statistical clustering properties of galaxies
by calculating the 2-point galaxy correlation function from the
Durham/UKST Galaxy Redshift Survey. This survey is magnitude limited to
\mbox{$b_{J} \sim 17$}, contains $\sim$2500 galaxies sampled at a rate
of one on three and surveys a $\sim$$4 \times 10^{6} (h^{-1}$Mpc$)^{3}$
volume of space. We have empirically determined the optimal method of
estimating the 2-point correlation function from just such a magnitude
limited survey. Applying our methods to this survey, we find
that our redshift space results agree well with those from previous
optical surveys. In particular, we confirm the previously claimed
detections of large scale power out to $\sim$40$h^{-1}$Mpc scales. We
compare with two common models of cosmological structure formation and
find that our 2-point correlation function has power significantly in
excess of the standard cold dark matter model in the 10-30$h^{-1}$Mpc
region. We therefore support the observational results of the APM
galaxy survey. Given that only the redshift space clustering can be
measured directly we use standard modelling methods and indirectly
estimate the real space 2-point correlation function. This real space
2-point correlation function has a lower amplitude than the redshift
space one but a steeper slope.
\end{abstract}

\begin{keywords}
galaxies: clusters -- galaxies: general -- cosmology: observations --
large-scale structure of Universe.
\end{keywords}

\section{Introduction}

Historically, the spatial 2-point correlation function, $\xi$, has
played a central role in the quantitative measurement of the strength
of galaxy clustering. It provides fundamental information about the
galaxy distribution in that sense that it is the Fourier transform
partner of the power spectrum of the density fluctuations. This
statistic is also both easy to compute, although quite laborious, and
easy to understand, with a direct probabilistic interpretation (e.g.
Peebles 1980).

The usual methods of estimating the spatial 2-point correlation
function are either from the deprojection of the angular correlation
function, $w(\theta)$, \cite{wtheta} or by direct estimation of the
observed galaxy distribution from redshift surveys \cite{dp83}. Both
methods have problems; the deprojection techniques are generally
unstable and require additional galaxy number-distance information,
while redshift surveys (by construction) have their galaxy distance
estimates distorted by the galaxy peculiar velocity field. Therefore,
they measure the real space correlation function after convolution with
this field.

The initial clustering results, redshift maps, etc. of the Durham/UKST
Galaxy Redshift Survey were summarized in the first paper of this
series \cite{mymnras}. In this paper we present a detailed analysis of
the 2-point correlation function clustering techniques and results from
this optically selected survey. We briefly describe our survey in
Section~\ref{redsursec}. The different methods of estimating the
2-point correlation function from a magnitude limited redshift survey
are described and tested in Section~\ref{2ptfnsec}. In
Section~\ref{xissec} we use the optimal method available to estimate
the galaxy 2-point correlation function from the Durham/UKST survey and
compare with the results from other galaxy redshift surveys and models
of structure formation. The projected 2-point correlation function is
described and estimated in Section~\ref{wvsec}. Finally, we summarize
our conclusions from this analysis in Section~\ref{concsec}.

\section{The Durham/UKST Galaxy Redshift Survey} \label{redsursec}

The Durham/UKST Galaxy Redshift Survey was constructed using the FLAIR
fibre optic system \cite{FLAIR} on the 1.2m UK Schmidt Telescope at
Siding Spring, Australia. This survey uses the astrometry and
photometry from the Edinburgh/Durham Southern Galaxy Catalogue (EDSGC;
Collins, Heydon-Dumbleton \& MacGillivray 1988; Collins, Nichol \&
Lumsden 1992) and was completed in 1995 after a 3-yr observing
programme. The survey itself covers a $\sim$$20^{\circ} \times
75^{\circ}$ area centered on the South Galactic Pole (60 UKST plates)
and is sparse sampled at a rate of one in three of the galaxies to
$b_{J} \simeq 17$ mag. The resulting survey contains $\sim$2500
redshifts, probes to a depth greater than $300h^{-1}$Mpc, with a median
depth of $\sim$$150h^{-1}$Mpc, and surveys a volume of space $\sim$$4
\times 10^{6} (h^{-1}$Mpc$)^{3}$.

The survey is $>$75 per cent complete to the nominal magnitude limit of
$b_{J} = 17.0$ mag. This incompleteness was mainly caused by poor
observing conditions, intrinsically low throughput fibres and other
various observational effects. In a comparison with $\sim$150 published
galaxy velocities (Peterson et al. 1986; Fairall \& Jones 1988;
Metcalfe et al. 1989; da Costa et al. 1991) our measured redshifts had
negligible offset and were accurate to $\pm 150$ kms$^{-1}$. The
scatter in the EDSGC magnitudes has been estimated at $\pm 0.22$ mags
\cite{nmb} for a sample of $\sim$100 galaxies. This scatter has been
confirmed by a preliminary analysis of a larger sample of high quality
CCD photometry. All of these observational details are discussed
further in a forthcoming data paper (Ratcliffe et al., in
preparation).

\section{Estimating the 2-Point Correlation Function from a Magnitude
limited Survey} \label{2ptfnsec}

In a volume limited, fair sample of the Universe (where the edge
effects of the galaxy survey can be neglected) an unbiased estimate of
the 2-point correlation function, $\xi(x)$, at separation $x$, is given
by
\begin{equation}
\xi(x) = \frac{DD(x)}{RR(x)}
\left(\frac{\bar{n}_{R}}{\bar{n}_{D}}\right)^{2} - 1 , \label{ddrreqn}
\end{equation}
where $DD(x)$ and $RR(x)$ are the data-data and random-random pair
counts at separation $x$ and $\bar{n}_{D}$ \& $\bar{n}_{R}$ are the
mean densities of the data (galaxy) \& random surveys, respectively.
However, for an apparent magnitude limited survey over a given fraction
of the sky things are not so simple. To estimate $\xi$ one has to deal
with a falling radial number density, how best to treat the edges of
the survey and the effects of being forced to calculate the mean
density internally from the survey itself. These problems manifest
themselves as the estimator we use to calculate $\xi$ and the
weighting we assign to each data/random point. We will take an
empirical approach to the solution of this problem and investigate the
different estimators and weightings equally.

\subsection{The Methods of Estimation} \label{ximethsec}

We will present results of the redshift space 2-point correlation
function, $\xi(s)$, where the redshift space separation between two
points $i$ and $j$ is given by
\begin{equation}
s = \sqrt{s_{i}^{2} + s_{j}^{2} - 2 s_{i} s_{j} \cos\theta} ,
\end{equation}
where $s_{i}$ and $s_{j}$ are the comoving redshift distances of the
two points separated by an angle $\theta$ on the sky (also see
Fig.~\ref{sigpifig}). Therefore, we have assumed a $q_{0} =
\frac{1}{2}$, $\Lambda = 0$ cosmology with comoving distances given by
\begin{equation}
s_{i} = \left( \frac{2c}{H_{0}} \right) \left[ 1 - \frac{1}{\sqrt{1 +
z_{i}}} \right] ,
\end{equation}
where $H_{0} = 100 h$kms$^{-1}$Mpc$^{-1}$ is the Hubble constant, $c$
is the velocity of light in kms$^{-1}$ and $z$ the observed redshift.

We calculate the radial selection function using standard methods
involving integrals over the galaxy luminosity function (e.g. Ratcliffe
et al. 1996b). Random points are then distributed radially within the
survey's angular limits with a probability proportional to the radial
selection function, volume element and completeness rate of the
survey. For the Durham/UKST survey this means distributing points
within each of the 60 UKST fields separately because not only are the
magnitude limits slightly different for each field (hence the radial
selection function is slightly different) but the completeness rates
are also slightly different. We have checked that this method of
distributing the random points does not cause any systematic biases
in~$\xi$, see Section~\ref{testximethsec}.

We then calculate the total number of data-data (DD), data-random (DR)
and random-random (RR) pair counts in the survey and bin according to
the pair separation of the points in question. We choose to bin our
counts in $0.1 dex$ bins of separation starting at $0.1h^{-1}$Mpc. We
calculate the 2-point correlation function using three different
estimators and two different weightings. The estimators investigated
here are the standard estimator (e.g. Peebles 1980)
\begin{equation}
\xi(x) = \frac{DD(x)}{DR(x)}
\left(\frac{\bar{n}_{R}}{\bar{n}_{D}}\right) - 1 , \label{dddreqn}
\end{equation}
the estimator proposed by Hamilton (1993)
\begin{equation}
\xi(x) = \frac{DD(x)RR(x)}{DR(x)^{2}} - 1 , \label{hameqn}
\end{equation}
and that of Landy \& Szalay (1993)
\begin{equation}
\xi(x) = \frac{DD(x) - 2DR(x) + RR(x)}{RR(x)} . \label{landseqn}
\end{equation}
The two weightings investigated here are a simple unit weighting
\begin{equation}
w(r) = 1 , \label{11eqn}
\end{equation}
and the so-called minimum variance weighting (Efstathiou 1988; Peebles
1973; Loveday et al. 1995)
\begin{equation}
w(r,x) = \frac{1}{1 + 4\pi n(r) J_{3}(x)} , \label{wweqn}
\end{equation}
where $n(r)$ is the radial number density and $J_{3}(x) = \int_{0}^{x}
\xi(y)y^{2}dy$ is the volume integral of the 2-point correlation
function out to a separation $x$. These estimators are essentially
Monte Carlo integrations over the spherical-shell shaped volumes of the
bins. These methods are particularly useful at the edges of the survey
where conventional integration techniques are impractical. In order to
reduce statistical fluctuations we use 25-50 times as many random
points as there are data points.

The standard estimator of equation~\ref{dddreqn} stood for many years
as the best estimate of $\xi$ from these types of survey, with the $RR
\rightarrow DR$ difference from equation~\ref{ddrreqn} giving a better
estimate for the Monte Carlo volume integration. However, this
estimator is sensitive to the error in the mean density \cite{ham}.
Estimators which are sensitive to the square of the error in the mean
density are those proposed by Hamilton (1993) and Landy \& Szalay
(1993). Also, while weighting each data/random point equally is the
simplest method, the pair count will be dominated by the structures in
the survey near the peak of the radial number density function. This
weighting essentially reduces the effective volume of the survey as
volumes are unequally sampled. To weight volumes equally one should
weight by the inverse of the radial selection function. Unfortunately,
such a weighting is dominated by the few galaxies at large distances,
where the selection function is small. Following on from work pioneered
by Peebles (1973), Efstathiou (1988) has proposed a weighting which
provides the mathematical minimum in the estimate of the variance of
$\xi$. This weighting turns out to be a happy medium
between equal pair weighting and equal volume weighting. To use
Efstathiou's (1988) weighing we need an estimate of $\xi$, namely the
quantity we are trying to calculate. This can be achieved via iteration
but in practice a $\left( r_{0}/r \right)^{\gamma}$ power law model for
$\xi$ suffices. We use the canonical values of $r_{0} = 5.0h^{-1}$Mpc
and $\gamma = 1.8$ (e.g. Peebles 1980). We include an upper limit of
$J_{3}^{max} = 5000h^{-3}$Mpc$^{3}$ in our weighting and find our
estimates of $\xi$ relatively insensitive to doubling/halving this
value.

\subsection{Testing the Methods} \label{testximethsec}

\begin{figure*}
\centering
\centerline{\epsfxsize=17.0truecm \figinsert{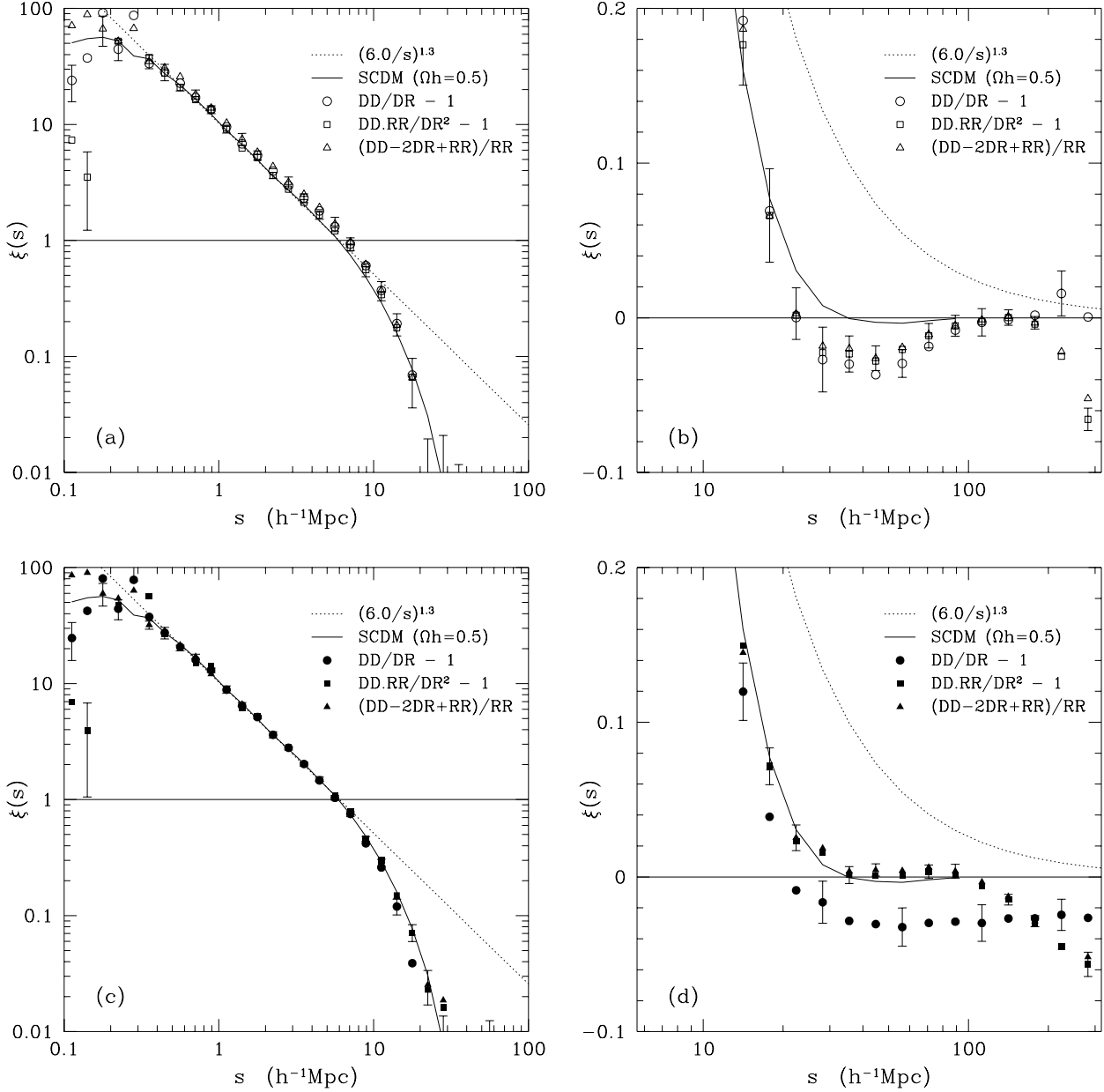}{0.0pt}}
\caption{Testing the methods of estimating the redshift space 2-point
correlation function, $\xi(s)$, from standard CDM mock catalogues which
mimic the Durham/UKST survey. On all of these plots open symbols denote
$w = 1$ unweighted estimates and closed symbols denote $ w = 1/(1 +
4\pi n(r) J_{3}(x))$ weighted estimates. Also, circular, square and
triangular symbols denote the estimators of
equations~\ref{dddreqn},~\ref{hameqn} and~\ref{landseqn},
respectively.  Figs.~(a) and~(c) are plotted on a log-log scale to
emphasize the small scale features, while Figs.~(b) and~(d) are plotted
on a log-linear scale to emphasize the large scale features. The dotted
line on each plot is the same simple power law model and can be used as
a reference point. The solid line is the redshift space 2-point
correlation function calculated directly from the $N$-body simulations
which are used to construct the mock catalogues. Error bars are the
$1\sigma$ scatter seen between the mock catalogues assuming each one
provides an independent estimate of $\xi$. To aid graphical clarity we
plot the alternate error bars of the three estimators.}
\label{scdmxifig}
\end{figure*}

\begin{figure*}
\centering
\centerline{\epsfxsize=17.0truecm \figinsert{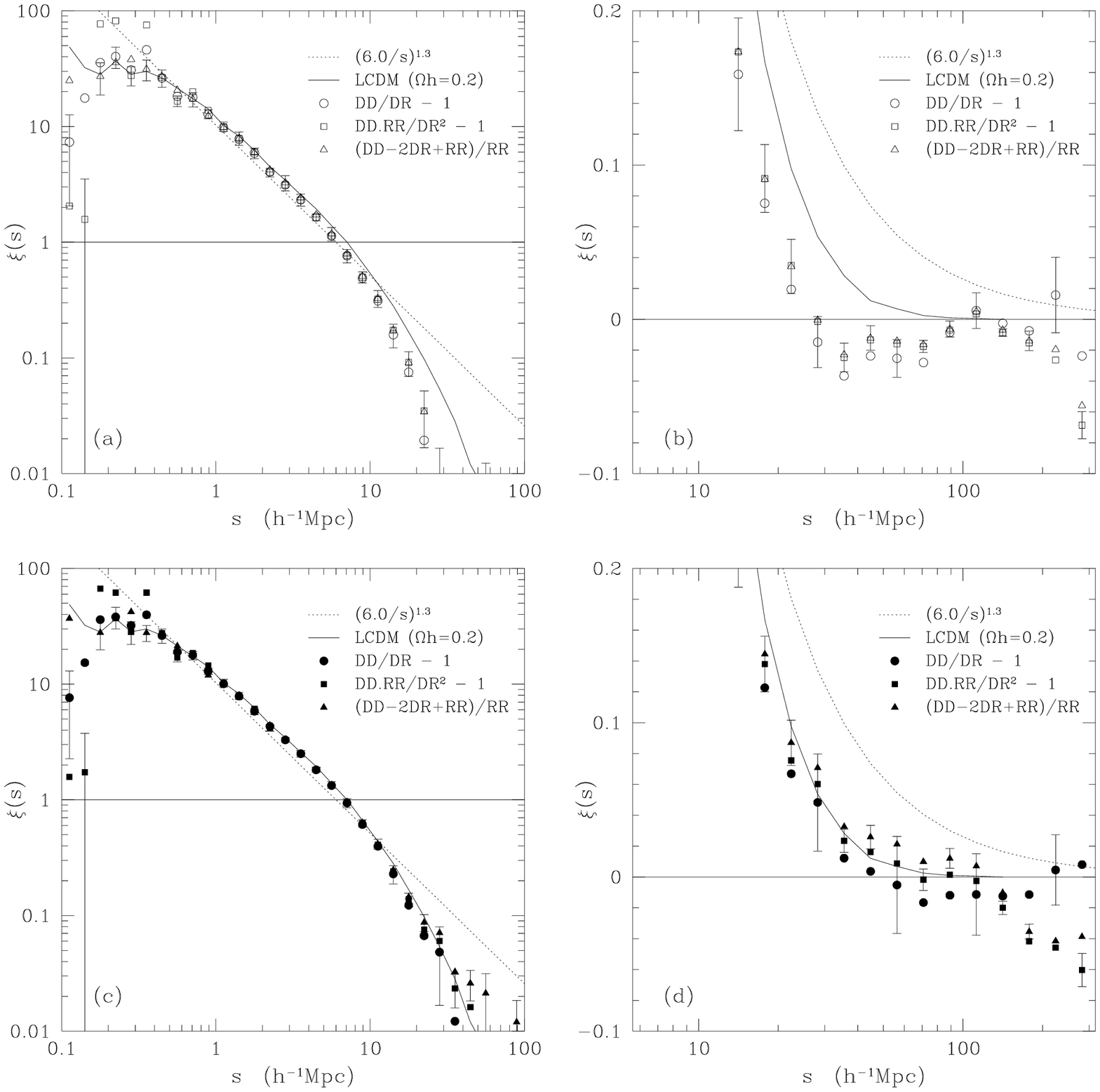}{0.0pt}}
\caption{The same as Fig.~\ref{scdmxifig} but for the mock catalogues
constructed from the low-$\Omega$ CDM model with a non-zero $\Lambda$
to ensure spatial flatness.} \label{lcdmxifig}
\end{figure*}

We have tested the reliability of the methods described in
Section~\ref{ximethsec} using mock catalogues of the Durham/UKST survey
drawn from two sets of cold dark matter (CDM) $N$-body simulations.
These mock catalogues were constructed in redshift space using the same
angular/radial selection functions and completeness rates as the actual
Durham/UKST survey. The CDM models used were (Efstathiou et al. 1985;
Gazta\~{n}aga \& Baugh 1995; Eke et al. 1996): standard CDM with
$\Omega h = 0.5$, $b = 1.6$ (SCDM); and CDM with $\Omega h = 0.2$, $b =
1$ and a cosmological constant ($\Lambda = 0.8$) to ensure a spatially
flat cosmology (LCDM). Each mock catalogue was selected in such a way
as to sample an independent volume of space from within the
simulation. Given the relative SCDM and LCDM comoving cube sizes (256
and 378$h^{-1}$Mpc), this implied that we could select a total of 18
SCDM mock catalogues from the 9 available SCDM simulations and 15 LCDM
mock catalogues from the 5 available LCDM simulations.

Figs.~\ref{scdmxifig} and~\ref{lcdmxifig} show the results of applying
the six different estimator and weighting combinations of
Section~\ref{ximethsec} to the SCDM and LCDM mock catalogues,
respectively. The circular, square and triangular symbols denote the
estimators of equations~\ref{dddreqn},~\ref{hameqn} and~\ref{landseqn},
respectively. Also, open symbols denote the {\it unweighted} estimates
of equation~\ref{11eqn} while closed symbols denote the {\it weighted}
estimates of equation~\ref{wweqn}. The dotted line on these plots is
the same simple power law model and as such can be used as a reference
point. The solid lines on these plots denotes the average of the actual
redshift space 2-point correlation function, $\xi(s)$, calculated
directly from the full N-body simulations (SCDM and LCDM, respectively).
Given that these are fully volume limited fair samples containing $N$
galaxies with a well defined mean density, $\bar{n}$, we use
equation~\ref{ddrreqn} and $RR = (4\pi/3) \bar{n}N (r_{outer}^{3} -
r_{inner}^{3})$, where $[r_{inner},r_{outer}]$ defines the extent of
the radial bin in question. The error bars shown are the $1\sigma$
standard deviation obtained from the observed scatter between the mock
catalogues. We have assumed that each mock catalogue provides a
statistically independent estimate of $\xi(s)$. Also, to aid graphical
clarity we only plot alternate error bars from the three estimators.

We also constructed a set of mock catalogues with constant completeness
rates in each field. The results obtained were almost identical and
therefore our method of distributing the random points does indeed
correct for the variable completeness rates of each field.  Another set
of mock catalogues were constructed in real space rather than redshift
space. Again, the results obtained were very similar and therefore the
conclusions of this section are independent of real and redshift space
effects.

From the SCDM mock catalogue results in Fig.~\ref{scdmxifig} we see
that, on small scales ($< 10h^{-1}$Mpc), all of the estimates can
reproduce the actual $\xi(s)$, although the weighted estimates are more
accurate and show less scatter. On large scales (10-100$h^{-1}$Mpc),
all of the unweighted estimates agree well but appear biased low by
$\sim 0.03$ in $\xi$. However, the weighted estimates trace the actual
$\xi(s)$ very well, except for the $DD/DR - 1$ estimator. These weighted
estimates also have smaller error bars than the unweighted ones, again
except for the $DD/DR - 1$ estimator. On very large ($> 100h^{-1}$Mpc) scales
we do not expect the mock catalogues to produce believable results
given the survey geometry involved.

We draw similar conclusions from the LCDM mock catalogue results of
Fig.~\ref{lcdmxifig}. This model has both a higher amplitude on small
scales and more power on large scales than the SCDM model. This time
the unweighted estimates are biased low by $\sim 0.08$ on large
(10-100$h^{-1}$Mpc) scales. Again, the weighted estimates trace the
actual $\xi(s)$ well on these scales. However, the weighted
$DD.RR/DR^{2} - 1$ estimator very accurately describes $\xi(s)$ on all
scales and also has the smallest error bars.

\subsection{Errors and Biases in the Estimates} \label{errbiassec}

The theoretical error in the 2-point correlation function on
large/linear scales has been estimated by Peebles (1973); see also
Kaiser (1986). Consider a wide bin containing $N_{p}$ data pairs in a
single radial shell with observed number density $n(r)$. Assuming that
$\xi$ is small ($\ll 1$) then the error in $\xi(x)$ is given by
\begin{equation}
\Delta\xi(x) = \frac{1 + 4\pi n(r) J_{3}(x)}{\sqrt{N_{p}}} .
\label{erreqn}
\end{equation}
This is essentially a $\sqrt{N}$ Poisson error modified for the effects
of clustering, which reduces the amount of independent information
available. We measure the maximum value of $4\pi J_{3}$ for the SCDM
and LCDM models to be $\sim$7000 and $17000h^{-3}$Mpc$^{3}$,
respectively. If we consider the survey as a whole then $N_{p} \simeq
n_{gal}^{2}$, where $n_{gal}$ is the total number of galaxies in the
survey. This implies a minimum theoretical error of $\Delta\xi \simeq
0.002$ and 0.007 in the SCDM and LCDM mock catalogues, respectively.
However, in studies of QSO clustering Shanks \& Boyle (1994) have
empirically shown that the error in equation~\ref{erreqn} only works
well on scales where $N_{p} < n_{gal}$. On scales where $N_{p} >
n_{gal}$ a more realistic error estimate is given by $\Delta\xi \simeq
1/\sqrt{n_{gal}}$. Given that we observe $N_{p} \simeq n_{gal} \simeq
2500$ on 5-10$h^{-1}$Mpc scales, we expect a minimum error of
$\Delta\xi \simeq 0.02$ on scales larger than this.

A possible bias in the estimation of $\xi$ is due to the fact that we
estimate both the mean density and the pair counts from the same
survey. This leads to a non-zero difference between the true $\xi$ of
an ensemble of surveys and the ensemble average of the $\xi$'s from
each survey. This is commonly known as the Integral Constraint (e.g.
Peebles 1980) and is given by
\begin{equation}
I_{c} \simeq \frac{1 + 4\pi n(r) J_{3}^{max}}{n_{gal}} ,
\end{equation}
which should be added to $\xi$ in an ensemble of surveys. One can
simplify this expression by assuming $1 \ll 4\pi n(r) J_{3}^{max}$ and
using $n(r) \simeq n_{gal}/V_{eff}$ to give
\begin{equation}
I_{c} \simeq \frac{4\pi J_{3}^{max}}{V_{eff}} ,
\end{equation}
where the effective volume sampled by the survey is given by
\begin{equation}
V_{eff} = \int_{V} f(r) dV ,
\end{equation}
and $f(r)$ is a function which reflects the weighting of the galaxies.
For example, if we weight pairs equally then $f$ is just the radial
selection function, while equal volume weighting implies that $f$ is
the inverse of the radial selection function. For a typical mock
catalogue we calculate $V_{eff} \sim 2 \times 10^{5}h^{-3}$Mpc$^{3}$
for equal pair weighting and $\sim 4 \times 10^{6}h^{-3}$Mpc$^{3}$ for
equal volume weighting. Recalling the maximum values of $4\pi J_{3}$
quoted previously we find that $I_{c} \simeq 0.035$ and 0.085 for the
SCDM and LCDM mock catalogues, respectively, when using equal pair
weighting and $I_{c} \simeq 0.002$ and 0.004 when using equal volume
weighting.

\subsection{Optimal Estimate}

\begin{figure}
\centering
\centerline{\epsfxsize=8.5truecm \figinsert{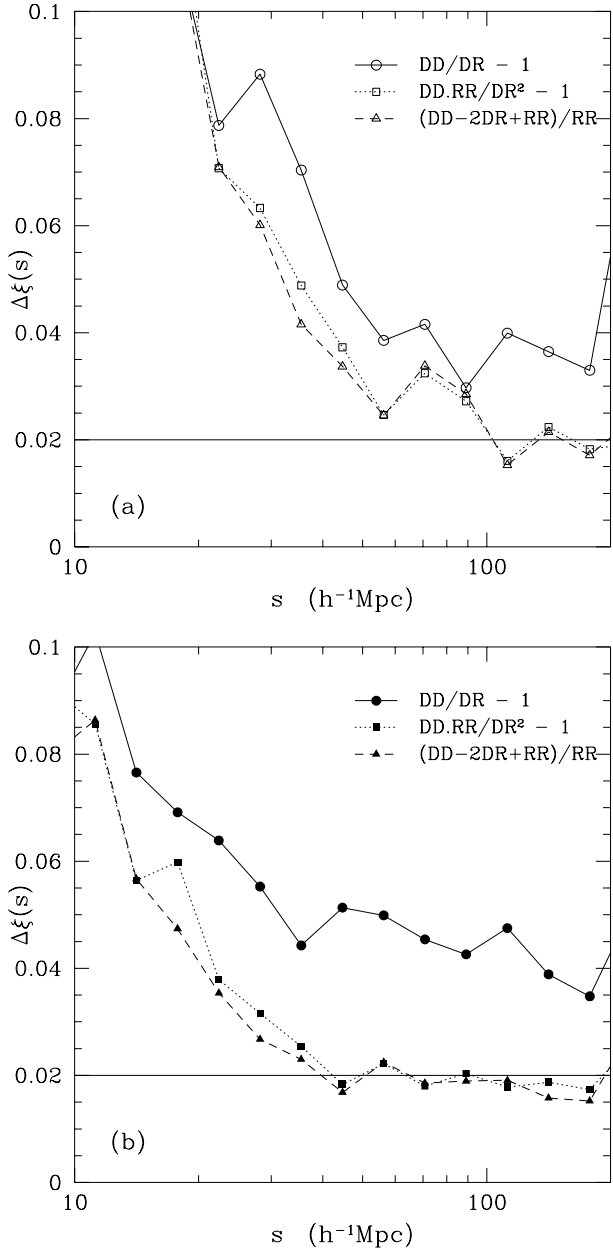}{0.0pt}}
\caption{An example of the error bars ($\Delta\xi$) on an individual
mock catalogue for the standard CDM model. Fig.~(a) shows the results
from the unweighted versions of the 3 estimators, while Fig.~(b) shows
the corresponding weighted estimates. These error bars appear to
asymptote to a value of $\Delta\xi \simeq 0.02$ on the larger scales,
in good agreement with our estimated minimum possible error bar. Very
similar results were found for the LCDM model.} \label{errfig}
\end{figure}

In Section~\ref{errbiassec} the realistic minimum error in an {\it
individual} mock catalogue was estimated to be $\Delta\xi \simeq 0.02$
on large scales, for both the SCDM and LCDM models. As an example, the
error bars on an individual SCDM mock catalogue are given in 
Figs.~\ref{errfig}(a) and~\ref{errfig}(b). These plots show that
this is a good estimate for the errors from both the weighted and
unweighted $DD.RR/DR^{2} - 1$ and $(DD-2DR+RR)/RR$ estimators and they
asymptote towards this value on large scales (10-100$h^{-1}$Mpc).
However, the most consistently small error bars are given by the
weighted $DD.RR/DR^{2} - 1$ and $(DD-2DR+RR)/RR$ estimators. 

We can also compare any systematic biases in the estimates of
Figs.~\ref{scdmxifig} and~\ref{lcdmxifig} with the predicted Integral
Constraint from Section~\ref{errbiassec}. We see that all of the
unweighted estimates suffer from a systematic bias which is in good
agreement with the predictions from the Integral Constraint: $\sim$0.03
compared with 0.035 for the SCDM mock catalogues; and $\sim$0.08
compared with $0.085$ for the LCDM mock catalogues. For the weighted
estimates there is no noticeable Integral Constraint for either set of
mock catalogues for the $DD.RR/DR^{2} - 1$ and $(DD-2DR+RR)/RR$
estimators. Again, this is in good agreement with the small value
predicted, namely $\leq 0.005$. We note that the weighted $DD.RR/DR^{2}
- 1$ estimate most accurately reproduces the actual $\xi$ of both the
SCDM and LCDM models on all scales.

Given the historical importance of the standard $DD/DR - 1$ estimator
we briefly discuss the results obtained from it. Empirically we observe
that the weighted estimate produces a larger error bar than the
unweighted estimate. This is in direct contradiction with the fact that
this weighting was constructed in order to produce the minimum variance
in $\xi$. This is only seen in the $DD/DR - 1$ estimates and therefore
could be due to the estimator itself. This is possibly related to the
fact that this estimator is sensitive to the error in the mean density
which is different from the other estimators which are sensitive to the
square of this error. Also, while the systematic bias in the unweighted
estimate can be explained by the Integral Constraint, the observed bias
in the weighted estimate remains unexplained. These results involving
the $DD/DR - 1$ estimator are in good agreement with a similar study of
pencil-beam surveys carried out by Fong, Hale-Sutton \& Shanks (1991).

To conclude this section we answer the question about which weighting
and estimator combination used on a mock catalogue optimally reproduces
the actual 2-point correlation function. We have found that all of our
estimates appear limited by a minimum error bar which comes directly
from the number of galaxies in the survey. Also, the unweighted
estimates of $\xi$ are all systematically biased low by an amount
predicted by the Integral Constraint. This is due to the fact that this
equal pair weighting reduces the effective volume of the survey.
Finally, we see that the weighting/estimator combination which most
accurately traces the actual $\xi$ and has the smallest error bars is
given by the $w = 1/(1 + 4\pi n(r) J_{3}(x))$ weighting of Efstathiou
(1988) and the $DD.RR/DR^{2} - 1$ estimator of Hamilton (1993). This is
what we call our optimal estimate of $\xi$ from a magnitude limited
redshift survey.

\section{The Redshift Space Galaxy 2-Point Correlation Function}
\label{xissec}

We estimate the redshift space 2-point correlation function, $\xi(s)$,
using the methods described in Section~\ref{2ptfnsec}. We use the
magnitude limits described in Ratcliffe et al. (1996b) which maximize
depth and minimize observational incompleteness in the survey. Using
these limits we have $\left< m_{lim} \right> = 16.86 \pm 0.25$ with an
average completeness rate of 75 per cent. Section~\ref{2ptfnsec} showed
that the methods of estimation were able to account for the effects of
having a slightly different magnitude limit and completeness rate in
each of the 60 UKST fields.

\subsection{Results from the Durham/UKST Survey}

\begin{figure}
\centering
\centerline{\epsfxsize=8.5truecm \figinsert{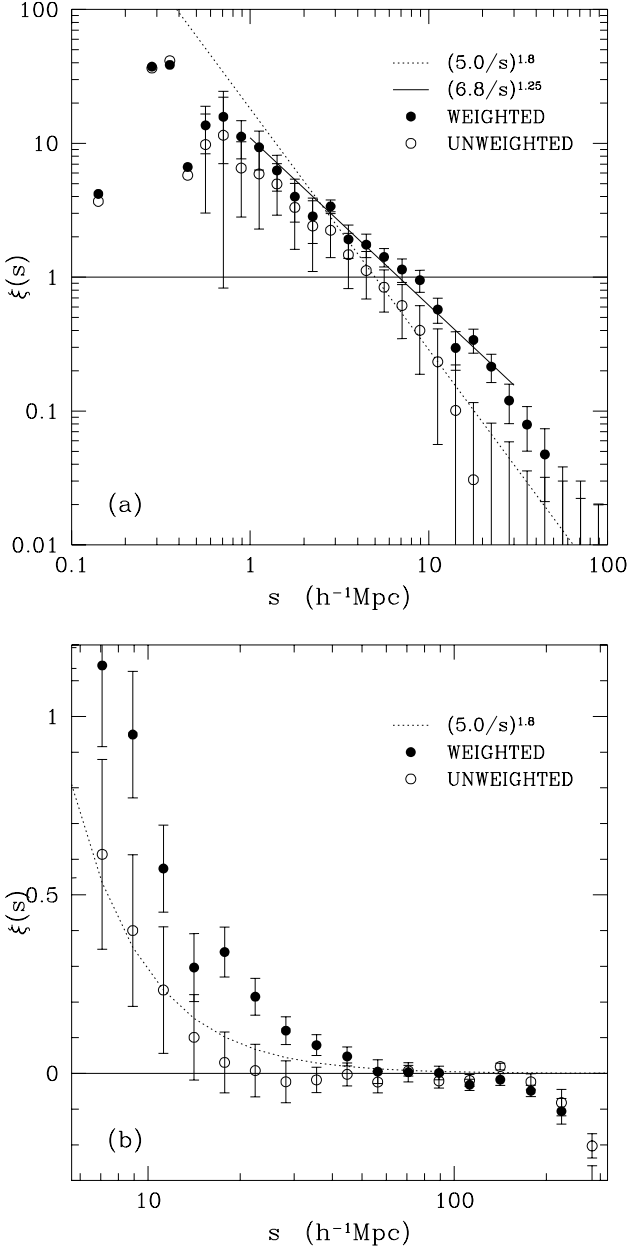}{0.0pt}}
\caption{Estimates of the redshift space 2-point correlation function,
$\xi(s)$, from the Durham/UKST Galaxy Redshift Survey using Hamilton's
(1993) estimator. Figs.~(a) and~(b) are plotted on log-log and
log-linear scales to emphasize the small and large scale features,
respectively. Open symbols denote the unweighted estimate and solid
symbols denote the weighted one. The dotted line shows the canonical
power law model, while the solid line shows the best fitting power law
model to the weighted $\xi(s)$ in the indicated range.}
\label{dur2ptfnfig}
\end{figure}

\begin{table}
\begin{center}
\caption{Comparison of the best fit redshift space 2-point correlation
function parameters from the Durham/UKST survey with recent galaxy
redshift survey results and also previous Durham ones.}
\label{xiscomptab}
\begin{tabular}{ccc}
Survey & {\large $s_{0}$} ($h^{-1}$Mpc) & {\large $\gamma$} \\
& & \\
Durham/UKST & $6.8 \pm 0.3$ & $1.25 \pm 0.06$ \\
APM-Stromlo & $5.9 \pm 0.3$ & $1.47 \pm 0.12$ \\
Las Campanas & $6.8 \pm 1.1$ & $1.70 \pm 0.11$ \\
DARS/SAAO & $6.5 \pm 0.5$ & ($1.8$) \\
\end{tabular}
\end{center}
\end{table}

Fig.~\ref{dur2ptfnfig} shows the results of applying these methods to
the Durham/UKST Galaxy Redshift Survey. We use Hamilton's (1993)
$DD.RR/DR^{2} - 1$ estimator but show the results from both the $w = 1$
unweighted and the $w = 1/(1 + 4\pi n(r) J_{3}(x))$ weighted estimates
for clarity. Figs.~\ref{dur2ptfnfig}(a) and~\ref{dur2ptfnfig}(b) are
plotted on log-log and log-linear scales to emphasize the small and
large scale features, respectively. Open symbols denote the unweighted
estimate and solid symbols denote the weighted one. The dotted line
shows the canonical power law model for $\xi$ of $\left( 5.0h^{-1}{\rm
Mpc}/s \right)^{1.8}$, while the solid line shows the best fitting
power law model to the weighted $\xi(s)$ in the 1-30$h^{-1}$Mpc range.
The error bars shown are the $1\sigma$ standard deviation on an
individual \mbox{low-$\Omega + \Lambda$} CDM mock catalogue (LCDM).
Obviously, these error bars use the same weighting/estimator
combination as the data points in question.

On all scales smaller than $\sim$100$h^{-1}$Mpc we see that the
unweighted estimate is systematically lower than the weighted one. We
have tested to see if this could be explained by any systematic errors
in the photometry, the method of incompleteness correction or the
errors in the measured redshifts and found a negative result. It
appears, quite simply, to be caused by the different weightings used.
Therefore, it is thought to be partially statistical and partially due
to the Integral Constraint. Indeed, using the value of $J_{3}^{max}$
estimated from the weighted $\xi(s)$ in Fig.~\ref{dur2ptfnfig} we find
that equal pair weighting could cause an Integral Constraint of
$\sim$0.25 in $\xi$. This is large enough to explain all of the
observed difference on $> 10h^{-1}$Mpc scales. Equal volume weighting
has an estimated Integral Constraint of $\sim$0.01 and is therefore not
a problem for the weighted estimate. We fit our power law model using a
minimum $\chi^{2}$ statistic and the best fit parameters are presented
in Table~\ref{xiscomptab}. This gave a $\chi^{2}$ of $\sim$10 for 13
degrees of freedom, which is an adequate fit. Errors on these
parameters come from the appropriate $\Delta\chi^{2}$ contour about
this minimum. However, given the correlated nature of these points, we
anticipate that our quoted errors are more than likely an
underestimate. This should be adequate for the simple comparison done
here.

Finally, given the results of Section~\ref{2ptfnsec}, we favour the
weighted $\xi(s)$ presented here as our best estimate of the redshift
space 2-point correlation function.

\subsection{Comparison with other Redshift Surveys}

Table~\ref{xiscomptab} also gives a comparison of the best fit power
law parameters of the $\xi(s)$ estimated from some recent optical
galaxy redshift surveys (Loveday et al. 1992, 1995; Tucker et al. 1996)
and also previous Durham ones (Shanks et al. 1983, 1989). We see that
the best fit redshift space correlation lengths, $s_{0}$, all agree
well with a value of $\sim$6.5$h^{-1}$Mpc. However, the slopes,
$\gamma$, all differ significantly given the quoted error bars. (Note
that the DARS/SAAO survey had $\gamma$ fixed at 1.8 during the
fitting.) Therefore, while the amplitude of $\xi(s)$ appears well
determined, there is considerable scatter in the value of the redshift
space slope from the currently available data sets.

\begin{figure}
\centering
\centerline{\epsfxsize=8.5truecm \figinsert{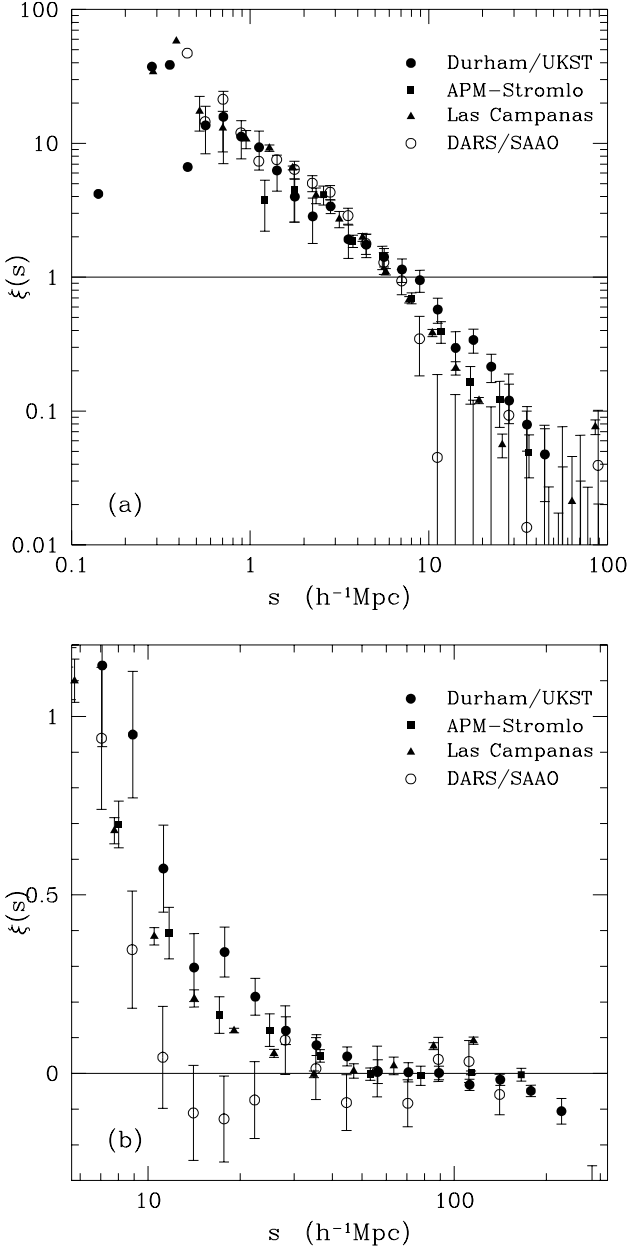}{0.0pt}}
\caption{Comparison of the Durham/UKST redshift space 2-point
correlation function, $\xi(s)$, with that from recent optical galaxy
redshift surveys (Loveday et al. 1992, 1995; Tucker et al. 1996) and
previous Durham surveys (Shanks et al. 1983, 1989). Figs.~(a) and~(b)
are plotted on log-log and log-linear scales to emphasize the small and
large scale features, respectively.} \label{xiscompfig}
\end{figure}

The results from these surveys are directly compared in
Figs.~\ref{xiscompfig}(a) and~\ref{xiscompfig}(b) where they are
plotted on log-log and log-linear scales to emphasize the small and
large scale features, respectively. The error bars shown on the
Durham/UKST estimate are again those from LCDM mock catalogues. On
small scales ($< 10h^{-1}$Mpc) we see that all of the estimates are
very consistent. On larger scales ($> 10h^{-1}$Mpc) we see that the new
Durham/UKST estimate agrees well with the previously claimed detections
of large scale power out to $\sim$40$h^{-1}$Mpc by the APM-Stromlo and
Las Campanas surveys. On even larger scales ($> 50h^{-1}$Mpc) all of
the surveys are consistent with zero. All of these $\xi$'s use the
estimator of Hamilton (1993) and the weighting of Efstathiou (1988),
apart from the previous Durham DARS/SAAO results. These authors used
the $DD/DR-1$ estimator with a $w=1$ weighting, but they did test against
the possibility that the integral constraint could be as large as
implied by the $w(\theta)$ found from the APM survey. Also, Fong et
al.  (1991) tested the effect of volume weighting the DARS/SAAO data
and found that the correlation function estimate only rose slightly;
they also found the increase in error from the combined used of volume
weighting and the $DD/DR-1$ estimator now reproduced in our analysis
here (see Fig.~\ref{xiscompfig}). Hence, we conclude that the reason that
the DARS/SAAO results are biased low is partly due to the use of equal
pair weighting but mainly due to statistical fluctuations in the
early redshift survey data.

Our conclusions from Table~\ref{xiscomptab} and Fig.~\ref{xiscompfig}
are that a simple, one power law model does not give a good fit to the
present data sets. However, the actual results from the different
surveys do in fact agree well on all scales on a qualitative level,
except for the DARS/SAAO results (which has large systematic errors).

\subsection{Comparison with Structure Formation Models}

\begin{figure}
\centering
\centerline{\epsfxsize=8.5truecm \figinsert{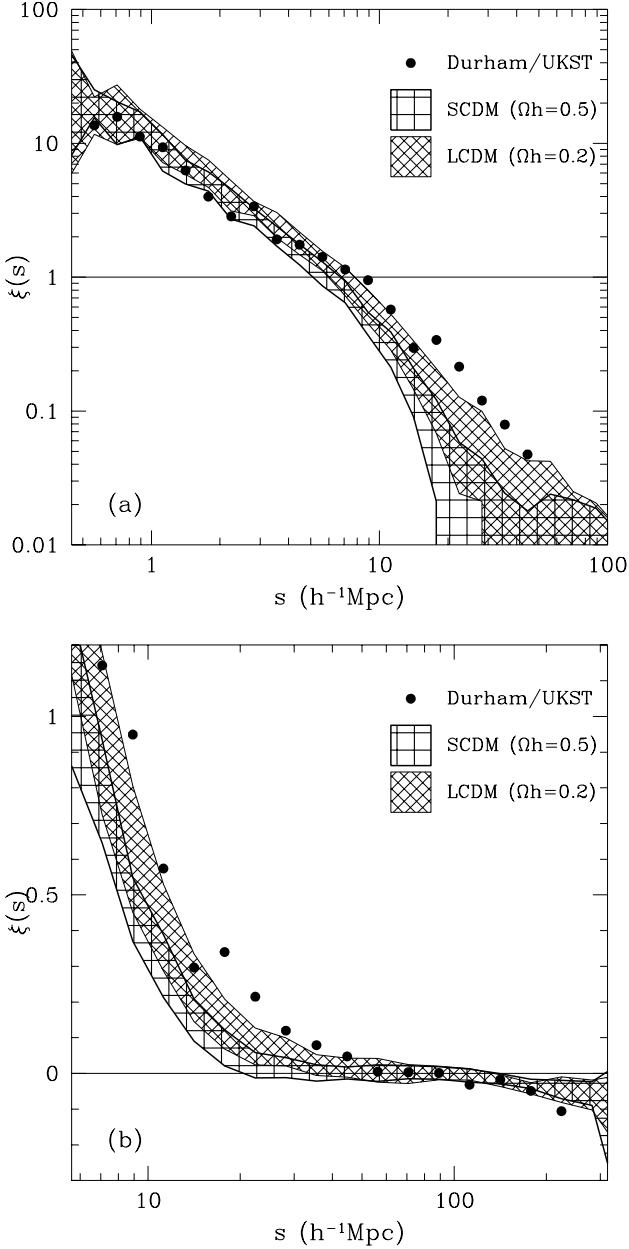}{0.0pt}}
\caption{Comparison of the Durham/UKST redshift space 2-point
correlation function, $\xi(s)$, with the results calculated from two
models of structure formation, namely the standard CDM model (SCDM) and
the \mbox{low-$\Omega + \Lambda$} CDM model (LCDM). Mock catalogues
which mimic the Durham/UKST survey were selected from the $N$-body
simulations of these models and the shaded areas denote the 68 per cent
confidence regions in $\xi(s)$ as estimated for an individual mock
catalogue. Figs.~(a) and~(b) are plotted on log-log and log-linear
scales to emphasize the small and large scale features, respectively.}
\label{xiscdmfig}
\end{figure}

We compare the redshift space 2-point correlation function from the
Durham/UKST survey with the predictions of two popular structure
formation models. The models we use are those from the cold dark matter
simulations of Section~\ref{2ptfnsec}, namely the standard CDM model
(SCDM) and the \mbox{low-$\Omega + \Lambda$} CDM model (LCDM).
Historically, the SCDM model has been the standard model of structure
formation for over 10 years \cite{scdm}, while the LCDM model is a
useful phenomenological model for recent large scale structure results
(e.g. Loveday et al. 1992; Baugh \& Efstathiou 1993). In
Figs.~\ref{xiscdmfig}(a) and~\ref{xiscdmfig}(b) we plot these results
on log-log and log-linear scales to emphasize the small and large scale
features, respectively. The shaded areas on Fig.~\ref{xiscdmfig} denote
the 68 per cent confidence region on an individual mock catalogue, see
Figs.~\ref{scdmxifig} and~\ref{lcdmxifig}. Given that the comparison
here is to see how often the CDM mock catalogues can reproduce the
Durham/UKST result (i.e. what is the scatter in the CDM estimates) we
do not plot error bars on the Durham/UKST estimate. For consistency,
all of the results presented in this figure were calculated using the
optimal weighting/estimator combination of Efstathiou (1988) and
Hamilton (1993).

On small scales ($< 10h^{-1}$Mpc) we see that both the CDM models agree
well with the results from the Durham/UKST survey. On large scales ($>
10h^{-1}$Mpc) the SCDM model shows no significant power above
$\sim$20$h^{-1}$Mpc whereas the LCDM model shows significant power out
to $\sim$30$h^{-1}$Mpc. Therefore, the Durham/UKST $\xi(s)$ has
significant power ($>3\sigma$) above and beyond the SCDM model up to
$\sim$40$h^{-1}$Mpc. While the LCDM model is more consistent with the
data, it also produces too little power in this region at the
1-2$\sigma$ level. This rejection of SCDM is consistent with the
findings from the APM galaxy survey (Maddox et al. 1990; Loveday et al.
1992) and the QDOT infrared redshift survey \cite{qdot}.

\section{The Projected Galaxy 2-Point Correlation Function}
\label{wvsec}

\begin{figure}
\centering
\centerline{\epsfxsize=7.5truecm \figinsert{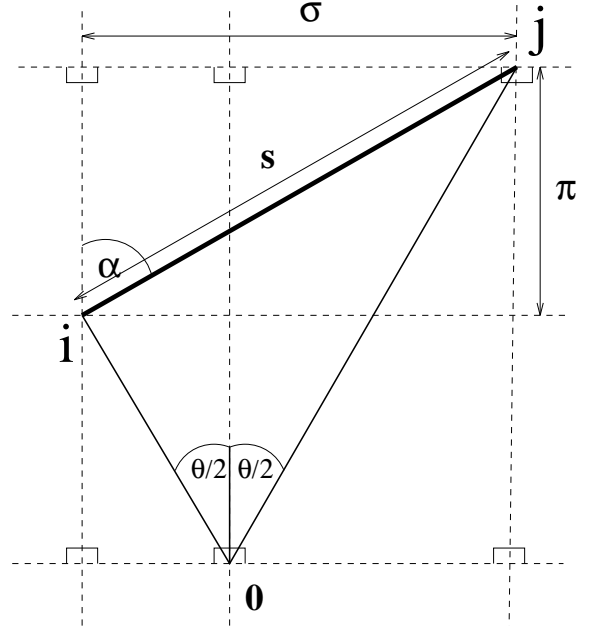}{0.0pt}}
\caption{Schematic diagram to show the definitions we use to calculate
the separations perpendicular ($\sigma$) and parallel ($\pi$) to the
line of sight of two points $i$ and $j$.} \label{sigpifig}
\end{figure}

Surveys which use measured redshifts to estimate distances have the
problem that the actual clustering pattern is imprinted with the galaxy
peculiar velocity field. Specifically, our distance estimates are
distorted by the non Hubble-flow component of the galaxy peculiar
velocity in the line of sight direction. Therefore, while our
fundamental interest (in clustering terms) is in the {\it real} space
2-point correlation function, only the {\it redshift} space 2-point
correlation function is directly observable from our survey. However,
it is possible to model the correlation function such that we can
estimate it as a simple real space power law.

We define the projected 2-point correlation function, $w_{v}(\sigma)$,
by (e.g. Peebles 1980)
\begin{eqnarray}
w_{v}(\sigma) & = & \int_{-\infty}^{\infty} \xi(\sigma,\pi) d\pi ,\\
& = & 2 \int_{0}^{\infty} \xi(\sigma,\pi) d\pi . \label{wvsigeqn}
\end{eqnarray}
where $\xi(\sigma,\pi)$ is our usual 2-point correlation function, but
calculated as a function of the separations perpendicular ($\sigma$)
and parallel ($\pi$) to the line of sight. The definitions of $\sigma$
and $\pi$ we use are schematically shown in Fig.~\ref{sigpifig}. We
found that our results do not depend significantly on the exact nature
of these definitions and even the small angle approximation gives
reasonably consistent results.

\subsection{Modelling the Projected Correlation Function}

The projected nature of equation~\ref{wvsigeqn} allows one to write
\begin{equation}
w_{v}(\sigma) = 2 \int_{0}^{\infty} \xi(\sqrt{\sigma^{2}+\pi^{2}})
d\pi,
\end{equation}
where $\xi(\sqrt{\sigma^{2}+\pi^{2}})$ is the real space 2-point
correlation function. Assuming a power law $\xi(r) = \left( r_{0}/r
\right)^{\gamma}$ with $r^{2} = \sigma^{2}+\pi^{2}$ and using the
definition of the Beta function gives
\begin{equation}
w_{v}(\sigma) = r_{0}^{\gamma} \left[\frac{
\Gamma\left(\frac{1}{2}\right) \Gamma\left(\frac{\gamma-1}{2}\right) }{
\Gamma\left(\frac{\gamma}{2}\right) }\right] \sigma^{(1-\gamma)} ,
\label{wvmodeqn}
\end{equation}
where $\Gamma(x)$ is the Gamma function and $\gamma > 1$ is assumed.
Therefore, we can fit for our measured $w_{v}(\sigma)$ to estimate a
power law model of $\xi(r)$.

\subsection{The Methods and Tests of the Methods}

\begin{figure}
\centering
\centerline{\epsfxsize=8.5truecm \figinsert{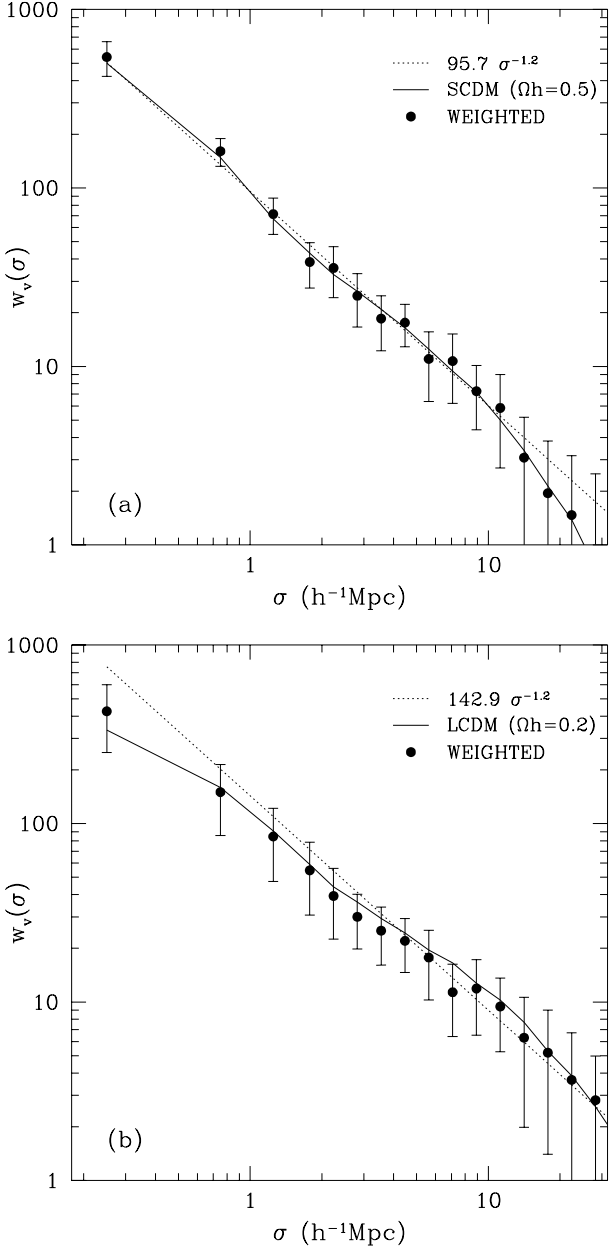}{0.0pt}}
\caption{Estimates of the projected 2-point correlation function for
(a) the SCDM model and (b) the LCDM model. The dotted line denotes the
power law model for $w_{v}(\sigma)$ predicted by
equation~\ref{wvmodeqn}. The solid line denotes the results of
estimating $w_{v}(\sigma)$ directly from the $N$-body simulations using
equation~\ref{wvcuteqn}. The solid points are the mean $w_{v}(\sigma)$
for the mock catalogues as estimated from equation~\ref{wvcuteqn}. The
error bars plotted are the $1\sigma$ standard deviation on an
individual mock catalogue.} \label{wvcdmfig}
\end{figure}

Our method for estimating $\xi(\sigma,\pi)$ is the same as in
Sections~\ref{2ptfnsec} and~\ref{xissec}, except that we now bin counts
in two variables instead of just one. The estimate of $\xi(\sigma,\pi)$
becomes noisy at very large scales and so we truncate the integral in
equation~\ref{wvsigeqn} at some upper limit, $\pi_{cut}$
\begin{equation}
w_{v}(\sigma) = 2 \int_{0}^{\pi_{cut}} \xi(\sigma,\pi) d\pi .
\label{wvcuteqn}
\end{equation}
In practise we use a $\pi_{cut}$ of $30h^{-1}$Mpc for all our
calculations and our results are insensitive to raising this value.
This integral is carried out using a simple mid-point integration
scheme which is quite adequate given the uncertainties in
$\xi(\sigma,\pi)$.

We test these methods by using the CDM $N$-body simulations and mock
catalogues of Section~\ref{2ptfnsec}. Firstly, although not shown here,
we have estimated the real space 2-point correlation function,
$\xi(r)$, directly from the $N$-body simulations in the same manner as
we estimated the actual redshift space 2-point correlation function for
Figs.~\ref{scdmxifig} and~\ref{lcdmxifig}. We find that the SCDM model
is approximately fit by a $r_{0} \simeq 5.0h^{-1}$Mpc, $\gamma \simeq
2.2$ power law out to $\sim$20$h^{-1}$Mpc scales. Similarly, the LCDM
model has approximate parameters of $r_{0} \simeq 6.0h^{-1}$Mpc,
$\gamma \simeq 2.2$ out to $\sim$30$h^{-1}$Mpc scales. These values of
$r_{0}$ and $\gamma$ are then used in equation~\ref{wvmodeqn} to
predict $w_{v}(\sigma)$ power laws of $\sim 95.7 \sigma^{-1.2}$ and
$142.9 \sigma^{-1.2}$ for the SCDM and LCDM models, respectively.
Secondly, we estimate the redshift space $\xi(\sigma,\pi)$ from each
$N$-body simulation directly and average to obtain the best estimate
possible for each CDM model. Using these two $\xi(\sigma,\pi)$'s we
then estimate $w_{v}(\sigma)$ from equation~\ref{wvcuteqn} for the SCDM
and LCDM models, respectively. Finally, we estimate $\xi(\sigma,\pi)$
from each mock catalogue using the optimal weighting/estimator
combination of Efstathiou (1988) and Hamilton (1993). Many estimates of
$w_{v}(\sigma)$ are obtained from equation~\ref{wvcuteqn} and then
averaged to produce the mean estimate from the SCDM and LCDM mock
catalogues, respectively.

We plot these three sets of results on Fig.~\ref{wvcdmfig}(a) for the
SCDM model and Fig.~\ref{wvcdmfig}(b) for the LCDM model. The dotted
line denotes the power law model for $w_{v}(\sigma)$ predicted by
equation~\ref{wvmodeqn}. The solid line denotes the results of
estimating $w_{v}(\sigma)$ directly from the $N$-body simulations using
equation~\ref{wvcuteqn}. The solid points are the mean $w_{v}(\sigma)$
for the mock catalogues as estimated from equation~\ref{wvcuteqn}. The
error bars on these points are the $1\sigma$ standard deviation on an
individual mock catalogue as calculated from the scatter between the
mock catalogues. Looking at Fig.~\ref{wvcdmfig}(a) we see that the
$w_{v}(\sigma)$ estimated from the SCDM mock catalogues (using the
optimal weighting/estimator of Section~\ref{2ptfnsec}) does accurately
reproduce the $w_{v}(\sigma)$ estimated directly from the SCDM $N$-body
simulations. Also, we see that the power law predictions of
equation~\ref{wvmodeqn} give good agreement with the estimated
$w_{v}(\sigma)$ out to $\sim$20$h^{-1}$Mpc scales, which is the scale
at which the power law approximation for the SCDM $\xi(r)$ was seen to
break down in the $N$-body simulations. We can make similar comments
regarding the LCDM results in Fig.~\ref{wvcdmfig}(b), namely that the
mock catalogues trace the expected $w_{v}(\sigma)$ and the predicted
power law model is a good approximation out to $\sim$30$h^{-1}$Mpc
scales, where the LCDM power law $\xi(r)$ breaks down.

To conclude these tests of the methods we state that the mock
catalogues do produce the expected projected 2-point correlation
function from the $N$-body simulations. Also, this method can
self-consistently reproduce the power law form of the real space
2-point correlation function from $\xi(\sigma,\pi)$ via the projected
2-point correlation function.

\subsection{Results from the Durham/UKST Survey}

\begin{figure}
\centering
\centerline{\epsfxsize=8.5truecm \figinsert{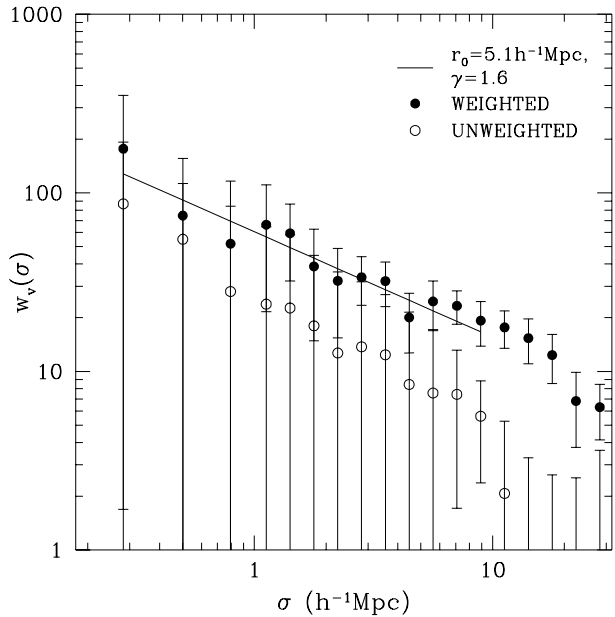}{0.0pt}}
\caption{Estimates of the projected 2-point correlation function,
$w_{v}(\sigma)$, from the Durham/UKST Galaxy Redshift Survey using
Hamilton's (1993) estimator. Open symbols denote the unweighted
estimate while solid symbols denote the weighted one. The solid line
shows the best fitting $\xi(r)$ power law model to the weighted
$w_{v}(\sigma)$ in the indicated range.} \label{wvdatafig}
\end{figure}

Fig.~\ref{wvdatafig} shows the results of applying these methods to the
Durham/UKST Galaxy Redshift Survey. We use Hamilton's (1993)
$DD.RR/DR^{2} - 1$ estimator to calculate $\xi(\sigma,\pi)$ but, for
clarity, show the results for both the unweighted estimate of
equation~\ref{11eqn} (open symbols) and the weighted estimate of
equation~\ref{wweqn} (solid symbols). The solid line shows the best
fitting power law model from equation~\ref{wvmodeqn} to the weighted
$w_{v}(\sigma)$ in the 0.25-10$h^{-1}$Mpc range. The error bars shown
are the $1\sigma$ standard deviation on an individual LCDM mock
catalogue. Obviously, these error bars use the same weighting/estimator
combination as the data points in question.

We see that the unweighted estimate is systematically lower than the
weighted one. This is a direct result of the weighted redshift space
2-point correlation function being higher than the unweighted one (see
Fig.~\ref{dur2ptfnfig}). The power law approximation of
equation~\ref{wvmodeqn} is fit using a minimum $\chi^{2}$ statistic and
the best fit parameters are presented in Table~\ref{xircomptab}. This
gave a $\chi^{2}$ of $\sim$8 for 12 degrees of freedom, which is an
adequate fit. Errors on these parameters come from the appropriate
$\Delta\chi^{2}$ contour about this minimum. However, given the
correlated nature of these points, we anticipate that our quoted errors
are more than likely an underestimate. Again, this should be adequate
for the simple comparison done here.

\begin{table}
\begin{center}
\caption{Comparison of the best fit real space 2-point correlation
function parameters from the Durham/UKST survey with recent galaxy
redshift survey results and also previous Durham ones.}
\label{xircomptab}
\begin{tabular}{ccc}
Survey & {\large $r_{0}$} ($h^{-1}$Mpc) & {\large $\gamma$} \\
& & \\
Durham/UKST & $5.1 \pm 0.3$ & $1.60 \pm 0.10$ \\
APM-Stromlo & $5.1 \pm 0.2$ & $1.71 \pm 0.05$ \\
Las Campanas & $5.0 \pm 0.14$ & $1.79 \pm 0.04$ \\
DARS/SAAO & $4.7 \pm 0.4$ & ($1.8$) \\
\end{tabular}
\end{center}
\end{table}

Table~\ref{xircomptab} also gives a comparison of the best fit $\xi(r)$
power law model parameters to the $w_{v}(\sigma)$ estimated from some
recent optical galaxy redshift surveys (Loveday et al. 1995; Lin et al.
1996) and also previous Durham ones (Bean et al. 1983; Hale-Sutton et
al. 1989). We see that the best fit real space correlation lengths,
$r_{0}$, all agree well with a value of $\sim$5.0$h^{-1}$Mpc. Also, the
slopes, $\gamma$, all agree quite well with a value of $\sim$1.75, bar
the Durham/UKST one which is \mbox{1-2$\sigma$} low. (Again the
DARS/SAAO survey had $\gamma$ fixed at 1.8 during the fitting.) We find
consistent results when comparing with the $r_{0} \simeq 4.5h^{-1}$Mpc
and $\gamma \simeq 1.7$ obtained by Baugh (1996) from numerically
inverting the APM angular correlation function, $w(\theta)$.

Our conclusion from Table~\ref{xiscomptab} is that a simple one power
law model gives both an adequate fit and consistent results from
present data sets.

\section{Conclusions} \label{concsec}

We have empirically determined the optimal method of estimating the
2-point correlation function from a magnitude limited galaxy redshift
survey. Our method used Monte Carlo techniques on mock catalogues drawn
from $N$-body simulations of cold dark matter structure formation
models. From the currently available choices of 2-point correlation
function estimator and weighting we find that both the minumum variance
and most accurate reproduction of the 2-point correlation function is
given by the estimator of Hamilton (1993) and the weighting of
Efstathiou (1988).

These techniques are then applied to the Durham/ UKST Galaxy Redshift
Survey and the redshift space 2-point correlation function is
calculated for this survey. We find that our results agree well with
those from other recent redshift surveys and confirm the previously
claimed detections of large scale power in the 10-40$h^{-1}$Mpc regime
(e.g. Loveday et al. 1992, 1995). A simple power law model is an
adequate fit to the data (although not particularly impressive) and has
redshift space parameters of correlation length, $r_{0} = 6.8 \pm 0.3
h^{-1}$Mpc, and slope, $\gamma = -1.25 \pm 0.06$. At small scales these
results agree with the results from previous Durham pencil-beam surveys
(Shanks et al. 1983, 1989). However, at large ($r>10h^{-1}Mpc$) scales
the older surveys suggested too litle power, mainly due to statistical
fluctuations, with some smaller contribution due to integral
constraint.

We compare our results with the predictions of two common models of
structure formation, namely the standard cold dark matter model,
$\Omega h = 0.5$ \& $b=1.6$ (SCDM), and a low density cold dark matter
model with a non-zero cosmological constant to ensure spatial flatness,
$\Omega h = 0.2$, $\Lambda = 0.8$ \& $b=1.0$ (LCDM). Our results agree
well with both of these models on small scales $< 10h^{-1}$Mpc but on
larger scales we find our results are $> 3 \sigma$ above and beyond the
SCDM model in the 10-40$h^{-1}$Mpc region. The LCDM model is more
consistent with our results but is still 1-2$\sigma$ low.

Given that our survey uses redshifts as distance estimates our measured
clustering statistics are distorted by the peculiar velocity field.
Using standard techniques (e.g. Peebles 1980) we calculate the
projected 2-point correlation function and use it to model the real
space 2-point correlation function. We find that a simple power law
model provides an adequate fit to the projected 2-point correlation
function from the Durham/UKST survey which implies real space
parameters of correlation length, $r_{0} = 5.1 \pm 0.3 h^{-1}$Mpc, and
slope, $\gamma = -1.6 \pm 0.1$. The differences seen in the real and
redshift space parameters is discussed in the next paper in this
series.

\section*{acknowledgments}

We are grateful to the staff at the UKST and AAO for their assistance
in the gathering of the observations. S.M. Cole, C.M. Baugh and V.R.
Eke are thanked for useful discussions and supplying the CDM
simulations. AR acknowledges the receipt of a PPARC Research
Studentship and PPARC are also thanked for allocating the observing
time via PATT and for the use of the STARLINK computer facilities.


\begin{thebibliography}{}

\bibitem[\protect\citename{Baugh }1996]{baugh} Baugh C.M., 1996,
MNRAS, 280, 267

\bibitem[\protect\citename{Baugh \& Efstathiou }1993]{baugh2} Baugh
C.M., Efstathiou G., 1993, MNRAS, 265, 145

\bibitem[\protect\citename{Bean et al. }1983]{dars3} Bean A.J.,
Efstathiou G., Ellis R.S., Peterson B.A., Shanks T., 1983, MNRAS, 205,
605

\bibitem[\protect\citename{Collins et al. }1988]{edsgc} Collins C.A.,
Heydon-Dumbleton N.H., MacGillivray H.T., 1988, MNRAS, 236, 7\textsc{p}

\bibitem[\protect\citename{Collins et al. }1992]{edsgc2} Collins C.A.,
Nichol R.C., Lumsden S.L., 1992, MNRAS, 254, 295

\bibitem[\protect\citename{da Costa et al. }1991]{ssrs} da Costa L.N.,
Pellegrini P.S., Davis M., Meiksin A., Sargent W.L., Tonry J.L., 1991,
ApJS, 75, 935

\bibitem[\protect\citename{e.g. Davis et al. }1985]{scdm} Davis M.,
Efstathiou, G., Frenk C.S., White S.D.M., 1985, ApJ, 292, 371

\bibitem[\protect\citename{e.g. Davis \& Peebles }1983]{dp83} Davis
M., Peebles P.J.E., 1983, ApJ, 267, 465

\bibitem[\protect\citename{Efstathiou }1988]{weight} Efstathiou G., in
Lawrence A., ed., 3rd {\it IRAS}, Conference, London, Comets to
Cosmology. Springer, Berlin, p. 312

\bibitem[\protect\citename{Efstathiou et al. }1985]{nbcode1} Efstathiou
G., Davis M., Frenk C.F., White S.D.M., 1985, ApJS, 57, 241

\bibitem[\protect\citename{Eke et al. }1996]{eke} Eke V.R., Cole S.M.,
Frenk C.S., Navarro J.F., 1996, MNRAS, 281, 703

\bibitem[\protect\citename{Fairall \& Jones }1988]{fj} Fairall A.P.,
Jones A., 1988, Publs. Dept. Astr. Cape Town, 10

\bibitem[\protect\citename{Fong et al. }1991]{fong} Fong R.,
Hale-Sutton D., Shanks T., 1991, in Blanchard A. et al. eds., 25th
Anniversary of the Cosmic Background Radiation Discovery. Editions
Frontieres, France, p.289

\bibitem[\protect\citename{Gazta\~{n}aga \& Baugh }1995]{gb}
Gazta\~{n}aga E. \& Baugh C.M., 1995, MNRAS, 273, 1\textsc{p}

\bibitem[\protect\citename{Hale-Sutton et al. }1989]{saao3} Hale-Sutton
D., Fong R., Metcalfe N., Shanks T., 1895, MNRAS, 237, 569

\bibitem[\protect\citename{Hamilton }1993]{ham} Hamilton A.J.S., 1993,
ApJ, 419, 19

\bibitem[\protect\citename{Kaiser }1986]{kaiser} Kaiser N., 1986,
MNRAS, 227, 1

\bibitem[\protect\citename{Landy \& Szalay }1993]{landy} Landy S.D.,
Szalay A.S., 1993, ApJ, 412, 64

\bibitem[\protect\citename{Limber }1954]{wtheta} Limber D.N., 1954,
ApJ, 119, 655

\bibitem[\protect\citename{Lin et al. }1996]{lin} Lin H., Kirchner
R.P., Tucker D.L., Shectman S.A., Landy S.D., Oemler A., Schechter
P.L., 1996, submitted to ApJ

\bibitem[\protect\citename{Loveday et al. }1992]{loveday} Loveday J.,
Efstathiou G., Peterson B.A., Maddox S.J., 1992, ApJ, 400, L43

\bibitem[\protect\citename{Loveday et al. }1995]{loveday2} Loveday J.,
Maddox S.J., Efstathiou G., Peterson B.A., 1995, ApJ, 442, 457

\bibitem[\protect\citename{Maddox et al. }1996]{maddox} Maddox et al.,
1990...

\bibitem[\protect\citename{Metcalfe et al. }1989]{saao} Metcalfe N.,
Fong R., Shanks T., Kilkenny D., 1989, MNRAS, 236, 207

\bibitem[\protect\citename{Metcalfe, Fong \& Shanks }1995]{nmb}
Metcalfe N., Fong R., Shanks T., 1995, MNRAS, 274, 769

\bibitem[\protect\citename{Parker \& Watson }1995]{FLAIR} Parker Q.A.,
Watson F.G., 1995, in Maddox S.J., Arag\'{o}n-Salamanca A., eds., 35th
Herstmonceux Conf. Cambridge, Wide Field Spectroscopy and the Distant
Universe. World Scientific, Singapore, p. 33

\bibitem[\protect\citename{Peebles }1973]{peeb73} Peebles P.J.E.,
1973, ApJ, 185, 413

\bibitem[\protect\citename{e.g. Peebles }1980]{peeb} Peebles P.J.E.,
1980, The Large-Scale Structure of the Universe, Princeton Univ.
Press, Princeton, NJ

\bibitem[\protect\citename{Peterson et al. }1986]{dars} Peterson B.A.,
Ellis R.S., Efstathiou G.P., Shanks T., Bean A.J., Fong R., Zen-Long
Z., 1986, MNRAS, 221, 233

\bibitem[\protect\citename{Ratcliffe et al. }1996a]{mymnras} Ratcliffe
A., Shanks T., Broadbent A., Parker Q.A., Watson F.G., Oates A.P., Fong
R., Collins C.A., 1996a, MNRAS, 281, L47

\bibitem[\protect\citename{Ratcliffe et al. }1996b]{mylfn} Ratcliffe
A., Shanks T., Parker Q.A., Fong R., 1996b, submitted to MNRAS

\bibitem[\protect\citename{Saunders et al. }1991]{qdot} Saunders W.,
Frenk C.S., Rowan-Robinson M., Efstathiou G., Lawrence A., Kaiser N.,
Ellis R.S., Crawford J., Xia x.-Y., Parry I., 1991, Nat, 349, 32

\bibitem[\protect\citename{Shanks et al. }1983]{dars2} Shanks T., Bean
A.J., Efstathiou G., Ellis R.S., Fong R., Peterson B.A., 1983, ApJ,
274, 529

\bibitem[\protect\citename{Shanks \& Boyle }1994]{shanks} Shanks T.,
Boyle B.J., 1994, MNRAS, 271, 753

\bibitem[\protect\citename{Shanks et al. }1989]{saao2} Shanks T.,
Hale-Sutton D., Fong R., Metcalfe N., 1989, MNRAS, 237, 589

\bibitem[\protect\citename{Tucker et al. }1996]{tucker} Tucker D.L.,
Oemler A.A., Shectman S.A., Kirshner R.P., Lin H., Landy S.D.,
Schechter P.L., 1996, preprint.

\end{thebibliography}
\end{document}